\documentclass[nofootinbib,showkeys,prd]{revtex4}
\usepackage{amsmath,amssymb,amsfonts,latexsym}
\usepackage[notcite,color,final]{showkeys}
\usepackage{amsmath,amsfonts,amssymb,amscd,graphicx}
\catcode`\@=11
\@addtoreset{equation}{section}
\catcode`\@=12
\def\de#1/de#2{\frac{\partial {#1}}{\partial {#2}}}
\begin{document}
\title{Weak Forces and Neutrino Oscillations under the standards of Hybrid Gravity with Torsion}
\author{Salvatore Capozziello$^{1,2}$, Luca Fabbri$^3$, Stefano Vignolo$^3$}
\affiliation{\it $^1$Dipartimento di Fisica, Università
di Napoli ``Federico II'', $^2$INFN Sez. di Napoli, Compl. Univ. di
Monte S. Angelo, Edificio G, Via Cinthia, I-80126, Napoli, Italy\\ $^3$ 
DIME Sez. Metodi e Modelli Matematici, Universit\`{a} di Genova,
Piazzale Kennedy, Pad. D - 16129 Genova, Italy}
\date{\today}
\begin{abstract}
We present a unifying approach where weak forces and neutrino oscillations are interpreted under the same standards of torsional hybrid gravity. This gravitational theory mixes metric and metric-affine formalism in presence of torsion and allows to derive an effective scalar field which gives rise to a running coupling for Dirac matter fields. In this picture, two phenomena occurring at different energy scales can be encompassed under the dynamics of such a single scalar field, which represents the further torsional and curvature degrees of freedom.
\end{abstract}
\maketitle
\section{Introduction}
It is the first time in the history of physics that the mismatch between theory and experiments takes so long to be quenched: in the standard model of particle physics, the theoretically predicted Higgs boson has been waiting for more than forty years its detection; in the standard model of cosmology, the experimentally observed phenomenon of neutrino oscillations is there since almost seventy years still lacking a model in which to fit. However, it may be that we are approaching the end of these two frustrating waits.

The standard model of particles seems today to be self-consistently fitted since the new boson recently observed has all features to make it a good candidate for the Higgs field; however, even if that were to be the case, that is even though what we have detected is the Higgs boson, it may well be that, at the present energies, we may recognize to be fundamental some field that at higher energies might be composite instead, and thus, although extensions based on no-Higgs models are likely to be soon ruled out, composite-Higgs models may still be part of the game; one of the first attempts in this direction was due to Weinberg himself (\cite{w}), modelling leptons in terms of a specific quartic potential for which leptons could bind to form a composite Higgs field, much in the same way in which quartic potentials are used to have the electrons binding together in Cooper pair and eventually forming Bose-Einstein condensates: in these models the Higgs will play the role of the Cooper couple, and its vacuum expectation value that of the Bose-Einstein condensate, the spontaneous being replaced by a dynamical breaking of the underlying symmetry.

For the issues of neutrino oscillations however the situation is more obscure, because the evidence is already there (at least at macroscopic level) and what is missing is the probe at fundamental level; the only attempt to describe self-consistently neutrino oscillation is that of Pontecorvo (\cite{p/1}, \cite{p/2} and \cite{g-p}), linking the oscillation length to the neutrino masses, but no neutrino mass has ever been seen directly. Further generalizations and extensions, relying on enlarged or alternative mechanisms, have been proposed, such as for instance those in \cite{a-b}, \cite{Ahluwalia:2011ea}, \cite{Ahluwalia:2009rp}, \cite{c-f} and \cite{p-r-w}.

From the General Relativity (GR) point of view, one characteristic that both attempts share is their neglecting torsional effects and so the full budget of geometric information seems not to be considered. In this perspective, it is natural to wonder what can be said if the torsional effects were kept in order to account for problems related to elementary particles.

In the standard model of particle physics, the presence of torsion gives rise to specific spinorial self-interactions within the Dirac equation that have the form of the potential Weinberg has been employing in \cite{w}; using these torsionally-induced interactions to model leptons does not only provide scalar but also axial-vector currents, so that the entire bosonic sector is supposed to be composite, and the uneven balance of chirality implies that there is no need to break any initial symmetry because there will be no symmetry that would ever be present in the first place \cite{Fabbri:2010ux}. With the same construction, neutrino oscillations are described not only in absence of neutrino masses but precisely because they are massless \cite{Fabbri:2010hz}. 

However, a closer look at these results also shows why there has been the general tendency to neglect torsion, that is torsion \emph{was} in fact negligible when taken at those energy scales.

Despite of the fact that torsional potentials may have intriguing effects in these cases, they nevertheless have no hope to ever be used so long as the torsional coupling constant is forced to be the Newton constant, because that would mean that torsional effects would only be relevant at the Planck scale; however, this is due to the fact that torsion entered the dynamical field equations beside curvature according to an action of the form $R$ (the Ricci scalar), written in terms of the connection containing both metric and torsion, and the fact that there is a single action for two independent fields should be seen as unsatisfactory: nevertheless, a more general action of the type $R+T^{2}$ (Ricci scalar and torsion squared) accounting for linear curvatures with implicit torsion plus explicit quadratic torsion terms endows the two independent fields with two independent coupling constants, so that although the gravitational one can be fixed at the Planck scale the torsional one is still undetermined, allowing torsion to be relevant at larger scales (\cite{Fabbri:2012qr} and \cite{Fabbri:2012yg}).

This may turn out to be quite intriguing, because if the new and still undetermined torsional coupling constant is chosen to have the value of the Fermi constant then the torsionally-inspired model of leptonic weak interactions with no symmetry breaking can in fact be equivalent to the Weinberg standard model of Leptons not only formally but also for its overall energy scale, that is not only in terms of the structure of the interaction but also for its strength \cite{Fabbri:2012zd}; also the neutrino oscillation length can be set to the value of the new torsional constant. But clearly these two physical situations are not likely to fit with one single value of the torsional coupling constant so more general theories should be considered than standard GR with torsion (see for instance \cite{fRgravity,fRgravity1,fRgravity2,fRgravity3,fRgravity4, revnoi, libro}).

Now, it is unlikely that torsion may enter with a coupling constant that can change from one physical situation to another; however, if the torsional coupling constant is allowed to display a running, then all these physical systems can be altogether interpolated in a unique framework. In this picture, one should search for mechanisms capable of producing a running coupling constant for torsion.

At a first sight this may be the hardest question to which we should reply, but fortunately this has already been considered: if the gravitational action is in more general forms than in standard GR, that is not confined to the linear action $R$ but is allowed to be a non-linear, generic function $f(R)$, then the torsional interactions do acquire a coupling running with the energy scale \cite{Capozziello:2012gw}. In this kind of theories, all effects for the above mentioned physical situations receive torsional contributions with a running coupling that makes them effective at different energy scales.

A possible critic that can be moved against these models is that the running couples to the energy scale algebraically, whereas a fully dynamical mechanism involving if possible also spin contributions would instead be preferable.

This too however is on the verge of receiving a proper treatment: in non-linear theories, a very recent approach has been pursued, the so-called Hybrid Gravity, consisting of taking into account the usual Einstein-Hilbert Lagrangian plus the $f(R)$ correction and insisting of employing both the purely metric and the metric-affine variational prescriptions, varying the Einstein-Hilbert Lagrangian with respect to the metric and the $f(R)$ term following the metric-affine technology (\cite{Harko:2011nh}, \cite{Capozziello:2012ny}); assuming torsion into this theory has the advantage that the torsional interactions have a running coupling that does not depend on external energy scales, being a dynamical field that can be found as the solution of some field equation, assigned within the theory itself. The running coupling is a scalar field described by scalar--like field equations, with an explicit spin dependence, given by the theory itself.

This approach would allow a straightforward solution for the above mentioned problems in what should be seen as the most physical way: as already mentioned, the theory is characterized by a scalar field undergoing scalar--like field equations where the presence of spin has a much larger impact, with the consequence that the coupling to fermions will turn out to have interesting dynamical effects. In this paper, we will study the consequences of this approach.

As an additional note, we first of all specify that throughout the whole paper we will always employ natural units, that is those for which $c=G=\hbar=1$: this choice is dictated by reasons of simplicity and it does not imply any loss of information; any lack of consistency in the mass dimension of different terms of the same equation is only apparent and the explicit dimension can be recovered at any moment with a straightforward dimensional analysis. For the rest of the paper, the organization is as it follows: in Sect.2, we deal with the problems of weak forces and neutrino oscillations, trying to set them under the same standards; Sect.3 is devoted to a summary on Hybrid Gravity where torsional degrees of freedom are considered, and in this view, weak forces for massive leptons and mixing oscillations among neutrinos are addressed; discussion and conclusions are drawn in Sect.4.
\section{Weak forces and neutrino oscillations under the same standards}
As we have discussed above, some of the most important open issues in physics, specifically weak forces and neutrino oscillations, could converge towards a common focusing point if a suitable theory of gravitational interaction is figured out in order to encompass all of them under the same standards.

The first problem is related to the standard model of leptons and weak interactions. This model, also known as Weinberg-Salam model, is based on the fusion between two essential ideas: $i)$ starting from an underlying symmetry giving rise to gauge fields in their massless configuration; $ii)$ having that symmetry spontaneously broken so to give masses to the weak mediators. The phenomenological aspects of the model are astonishing: the lepton sector is known with the precision of one part in a million, the weak sector, one part in a thousand. However, in order to obtain the symmetry breaking, an additional ingredient is needed, the Higgs field and all its potentials, which also contribute to dynamics; although missing for forty years, recently a boson has been found at CERN, for which the indications that it is the Higgs field itself are extremely strong. This however would not end the research about the Higgs sector.

In fact, even though that particle is practically identified with the Higgs scalar, nothing tell us that at higher-energies the Higgs field would not turn out to be something different from what the standard model foresaw: that is the Higgs field might turn out not to be fundamental; it might be composite. In fact, there are also many theoretical reasons for wanting a composite Higgs field, as first discussed by Weinberg himself; since then, many proposals have been put forward, the most important of which being Technicolor and Fermion Condensates. Both approaches are based on a model of Higgs field as fermion composite field bound together, in the first case, by the technicolor gauge bosons, and, in the second case, by $4$-fermion interaction \cite{b-h-l}; because technicolor bosons have yet to be observed, we think it is wiser to focus attention to the $4$-fermion interaction. 

In the simplest situation given by a $4$-fermion potential for two lepton fields generating the leptonic field equations
\begin{eqnarray}
&i\gamma^{\mu}\partial_{\mu}e
-\frac{1}{\varphi}\overline{e}\gamma_{\mu}e\gamma^{\mu}e
-\frac{1}{\varphi}\overline{\nu}\gamma_{\mu}\nu\gamma^{\mu}\gamma e-me=0
\label{e}\\
&i\gamma^{\mu}\partial_{\mu}\nu
-\frac{1}{\varphi}\overline{e}\gamma_{\mu}\gamma e\gamma^{\mu}\nu=0
\label{n}
\end{eqnarray}
as field equations that has to hold in general. Notice that the interaction induces an effective coupling between each chiral projection with all other opposite-chirality projections: in particular, what this means is that single-handed fields like the neutrinos do not have self-interactions; but more in general any field, whether single-handed or double-handed, has mutual interactions with all other fields. Thus, the dynamics is as it follows: when any given lepton propagates close to another lepton enough for the latter to have relevant contribution to the dynamics of the former, then an effective interaction, mediated by torsion, occurs. This is true for the electron-neutrino interaction, giving rise to an effective weak force. In this case the interaction does not mix electron and neutrino because they have different masses.

Another problem still related to particle physics but at much larger scales is that of the neutrino oscillations. Neutrino oscillations have definitively been confirmed quite recently at the SNO experiment, and for them to be explained in terms of simpler concepts, we have the possibility to link their oscillation to the presence of neutrino masses through the Pontecorvo model: in this model, the effective mixing takes place because of the neutrino masses. The model seems to yield the correct oscillation length, although fine-tunings must at some point be taken into account.

For the Pontecorvo model to work, neutrinos must be massive, but no measurement of neutrino masses has ever been performed directly: what this implies is that any alternative model in which neutrinos are massless, but nevertheless able to oscillate, would work just as well. In order to produce neutrino oscillations directly, we would then need an effective neutrino interaction, which could exist in complete analogy with the case mentioned above for the $4$-fermion interaction. 

This will be given in the case of an even simpler situation in which the $4$-fermion potential is given for two lepton fields that are both neutrino fields generating the neutrino field equations
\begin{eqnarray}
&i\gamma^{\mu}\partial_{\mu}\nu_{1}
+\frac{1}{\phi}\bar{\nu}_{2}\gamma_{\mu}\nu_{2}\gamma^{\mu}\nu_{1}=0
\label{n1}\\
&i\gamma^{\mu}\partial_{\mu}\nu_{2}
+\frac{1}{\phi}\bar{\nu}_{1}\gamma_{\mu}\nu_{1}\gamma^{\mu}\nu_{2}=0
\label{n2}
\end{eqnarray}
in general, and of course a similar equation would hold for any other couple of neutrinos no matter the family. In this case, the interaction induces a coupling even though both fields are single-handed: this will also imply that the effective interaction can mix the neutrino flavours since they are both massless. As a consequence, neutrino-number, and more in general, lepton-number, are violated as part of the mechanism, as expected.

Now the question is whether there is the possibility to get the field equations (\ref{e}-\ref{n}) and (\ref{n1}-\ref{n2}) both at once.

However, even supposing that this result could be achieved, there is a problem with the construction. In all equations above, the corrections came into the picture with the correspondent coupling constants, whose values are unknown, but it is unlikely that the same coupling constant might fit both different physical situations with the same value: in the case of leptonic weak forces the constant must be chosen to be the Fermi constant; in the case of neutrino oscillation that constant has to be related to the neutrino oscillation length. It is hardly the case that the Fermi constant could also give rise to the correct length: so a question that now arises is whether it could be possible to have a theory in which this problem is settled up to astrophysical scales (see \cite{annalen} and \cite{arturo} for details). Clearly, the easiest solution would be given by a theory where the two coupling constants are two different values, taken at two different energy scales, of a single function of the energy, which therefore can be thought as a running coupling.

So the question is now if it possible to get equations (\ref{e}-\ref{n}) and (\ref{n1}-\ref{n2}) with two couplings that are interpolated by a single running coupling.

A running coupling, where the running goes with the energy scale, is a feature of the so-called $f(R)$-theories of gravity where the gravitational action is given by a generic function of the Ricci scalar $R$ \cite{libro}; in $f(R)$-gravity, when the field equations are obtained, they can be worked out to be written as those we would have had in the standard Einstein gravity plus corrections of essentially two types: $i)$ one is the presence of additional terms in the form of a energy density tensor for the scalar field $\phi=f'(R)$, $ii)$ the other is the presence of factors $\frac{1}{\phi}$ constituting some sort of scaling. This is encouraging, but there is a problem: in $f(R)$-gravity with torsion the scaling depends algebraically on the energy, whereas it would be physically preferable to have this function assigned dynamically. The easiest way in which this can be done consists in having this function, which is a scalar field obeying scalar-like field equations. In the next section, we will discuss a possible realization of this issue.
\section{An encompassing approach under Hybrid Gravity with Torsion}
I order to match the issues of weak forces and neutrino oscillations under the same standards, we need a running coupling constant scaling with energies which should give rise to interactions for Dirac fermionic matter fields. The only interaction capable of connecting such so different scales is gravity; besides, due to the presence of massive spin particles, the gravitational theory has to contain curvature and torsional degrees of freedom: the so-called Hybrid Gravity endowed with torsion could be the approach to unify the above different problems. Such a theory takes into account the standard GR, in metric formalism, corrected by $f(R)$ gravity, assumed {\it a la Palatini}, in the same effective action (\cite{Harko:2011nh} and \cite{Capozziello:2012ny}). 

In such a way, the correct weak field limit, reproducing the Solar System constraints, is achieved and features of $f(R)$-gravity, useful for cosmology and astrophysics, can be considered: the main point of such an approach is that the theory can be expressed in a dynamically equivalent scalar-tensor form where a running coupling is naturally achieved. Due to this characteristic the above problems could be encompassed under the same standards.

To summarize the main features of hybrid gravity with torsion, we begin by recalling the notations and conventions we will adopt, starting from the basic quantities given by the metric $g_{ij}$ and the metric-compatible connection ${\Gamma}^{\;\;\;h}_{ij}$ with torsion
\begin{equation}
\label{torsion}
T_{ij}^{\;\;\;h}
=\Gamma_{ij}^{\;\;\;h}-\Gamma_{ji}^{\;\;\;h}
\end{equation}
for which the connection can be decomposed according to
\begin{equation}
\label{connectiondec}
\Gamma_{ij}^{\;\;\;h}=\tilde{\Gamma}_{ij}^{\;\;\;h}-K_{ij}^{\;\;\;h}
\end{equation}
in terms $\tilde{\Gamma}_{ij}^{\;\;\;h}$ called Levi-Civita connection and where 
\begin{equation}
\label{contorsion}
K_{ij}^{\;\;\;h}
=\frac{1}{2}\left(-T_{ij}^{\;\;\;h}+T_{j\;\;\;i}^{\;\;h}-T^{h}_{\;\;ij}\right)
\end{equation}
is called the contorsion tensor; from the connection we can define the Riemann tensor
\begin{equation}
\label{riemann}
R^{h}_{\;\;kij}
=\partial_i\Gamma_{jk}^{\;\;\;h} - \partial_j\Gamma_{ik}^{\;\;\;h} +
\Gamma_{ip}^{\;\;\;h}\Gamma_{jk}^{\;\;\;p}-\Gamma_{jp}^{\;\;\;h}\Gamma_{ik}^{\;\;\;p}
\end{equation}
decomposable according to
\begin{equation}
\label{riemanndec}
R^{h}_{\;\;kij}=\tilde{R}^{h}_{\;\;kij}-
\tilde{\nabla}_iK_{jk}^{\;\;\;h}+\tilde{\nabla}_jK_{ik}^{\;\;\;h} +
K_{ip}^{\;\;\;h}K_{jk}^{\;\;\;p}-K_{jp}^{\;\;\;h}K_{ik}^{\;\;\;p}
\end{equation}
in terms of the Riemann curvature of the Levi--Civita connection $\tilde{R}_{ij}$ and the Levi--Civita derivatives $\tilde{\nabla}_h$ of the contorsional contributions.

With these quantities, the fundamental action we will consider is given by \cite{Harko:2011nh}
\begin{equation}
\label{action}
S = \int \left[\tilde{R} + f(R) \right]\sqrt{-g}d^4x + S_m
\end{equation}
where, in addition to the standard Einstein-Hilbert term $\sqrt{-g}\tilde{R}$, there is an extra term depending on both metric and connection or equivalently metric and torsion, and where $S_m$ is the material action: as for this action, two formulations can be used, either the metric-affine, varying with respect to metric and connection (or torsion), or the tetrad-affine, varying with respect to tetrads and spin-connection. When the matter Lagrangian is independent of the dynamical connection, both are equivalent, but otherwise the tetrad-affine framework is in general necessary when dealing with spinors, as the fermionic Dirac field.

The variations of (\ref{action}) can be worked out in a similar way as for $f(R)$-gravity with torsion (\cite{CCSV1}, \cite{CCSV2}, \cite{CV4}), obtaining field equations of the form 
\begin{subequations}
\label{fe}
\begin{equation}
\label{feenergy}
\tilde{G}_{ij} + f'(R)R_{ij} -\frac{1}{2}f(R)g_{ij} = \Sigma_{ij}
\end{equation}
\begin{equation}
\label{fespin}
f'(R)T_{ij}^{\;\;\;h}=\frac{1}{2}\left(\partial_p\/{f'(R)} + S_{pq}^{\;\;\;q}\right)
\left(\delta^{p}_{j}\delta^{h}_{i}-\delta^{p}_{i}\delta^{h}_{j}\right)+S_{ij}^{\;\;\;h}
\end{equation}
\end{subequations} 
where $\tilde{G}_{ij}$ denotes the Einstein tensor associated with the Levi--Civita connection while $G_{ij}$ would be the Einstein tensor of the full connection, $\Sigma_{ij}$ and $S_{ij}^{\;\;\;h}$ represent the energy and the spin density tensors deriving from variations of the matter action $S_m$ as it is known: from the trace of equations (\ref{feenergy}) we derive the relation
\begin{equation}
\label{feenergytrace}
f'(R)R-2f(R)=\Sigma+\tilde{R}
\end{equation}
between the scalar curvature $R$ and the variable $X:=\tilde{R}+\Sigma$ which measures how much the theory deviates from the trace equation $\tilde R+\Sigma=0$ defining Einsteinian gravity \cite{Capozziello:2012ny}; from field equation (\ref{fespin}) we derive the two following relations
\begin{subequations}
\begin{equation}
\label{fespintrace}
-\frac{1}{3}\left[2f'(R)T_{ih}^{\;\;\;h}+S_{iq}^{\;\;\;q}\right]=\partial_if'(R)
\end{equation}
\begin{equation}
\label{fespinaxial}
\varepsilon^{aijh}T_{ijh}=\frac{1}{f'(R)}\varepsilon^{aijh}S_{ijh}
\end{equation}
\end{subequations}
showing that the trace of torsion and spin encodes the information about the non-linear character of the $f(R)$ while the completely antisymmetric parts of torsion and spin are directly correlated, a fact that will be important when dealing with the completely antisymmetric spin of Dirac matter fields. We rewrite (\ref{fe}) as
\begin{equation}
\label{2.7abis}
\tilde{G}_{ij} + f'(R)G_{ij} = \Sigma_{ij} - \frac{1}{2}\left[ f'(R)R -f(R) \right]\/g_{ij} 
\end{equation}
which will be important in the following.

Also, it is worth mentioning the conservation laws
\begin{subequations}
\label{2.11}
\begin{equation}
\label{2.11a}
\nabla_i\Sigma^{ij} + T_i\Sigma^{ij} - \Sigma_{pq}T^{jpq} -\frac{1}{2}S_{pqr}R^{pqrj} = 0
\end{equation}
\begin{equation}\label{2.11b}
\nabla_h\/S^{ijh} + T_h\/S^{ijh} + \Sigma^{ij} - \Sigma^{ji} = 0 
\end{equation}
\end{subequations}
which have to be satisfied by the energy and spin density tensors once the matter field equations are assigned (an explicit proof of these conservation laws (\ref{2.11}) for general Lagrangians depending on tetrad and spin-connection can be found in \cite{Poplawski}, where the invariance of the Lagrangian under diffeomorphisms and Lorentz transformations is used).

Having established the general form of such field equations, we may now proceed to prove the equivalence of hybrid gravity with torsion and scalar-tensor theories. 
To this purpose, let us consider the metric-affine action
\begin{equation}\label{3.1}
S = \int \left[ \tilde{R} + \varphi\/R -V(\varphi) \right]\sqrt{-g}d^4\/x + S_m
\end{equation}
which represents a metric-affine (with torsion) scalar-tensor theory, with Brans-Dicke parameter $w=0$ and potential $V(\varphi)$ for the scalar field $\varphi$: by varying the action (\ref{3.1}) with respect to the scalar $\varphi$, the metric and the connection, we get the field equations
\begin{subequations}
\label{3.2}
\begin{equation}
\label{3.2a}
\tilde{G}_{ij} + \varphi\/G_{ij} = \Sigma_{ij} -\frac{1}{2}V(\varphi)g_{ij}
\end{equation}
\begin{equation}\label{3.2b}
\varphi T_{ij}^{\;\;\;h}=\frac{1}{2}\left(\partial_p\/\varphi + S_{pq}^{\;\;\;q}\right)
\left(\delta^{p}_{j}\delta^{h}_{i}-\delta^{p}_{i}\delta^{h}_{j}\right)+S_{ij}^{\;\;\;h}
\end{equation}
\begin{equation}\label{3.2c}
R = \frac{dV}{d\varphi}
\end{equation}
\end{subequations}
Denoting by $F(R):=f'(R)$ and supposing the function $F(R)$ invertible, the potential $V(\varphi)$ can be chosen as
\begin{equation}\label{3.3}
V(\varphi):=\varphi\/F^{-1}(\varphi) - f(F^{-1}(\varphi))
\end{equation}
Under this assumption, it is straightforward to see that eq. (\ref{3.2c}) amounts to the identity
\begin{equation}
\label{3.4}
\varphi = F(R)=f'(R)
\end{equation}
In such a circumstance, eqs. (\ref{3.2a}) and (\ref{3.2b}) are then identical to (\ref{feenergytrace}) and (\ref{fespin}), showing the dynamical equivalence of the actions (\ref{action}) and (\ref{3.1}).

As a final step, it is possible to see that from eq. (\ref{3.2b}) we can obtain the following representation for the contorsion tensor
\begin{equation}\label{2.8}
K_{ij}^{\;\;\;h}= \hat{K}_{ij}^{\;\;\;h} + \hat{S}_{ij}^{\;\;\;h}
\end{equation}
where we have defined
\begin{subequations}\label{2.9}
\begin{equation}\label{2.9a}
\hat{S}_{ij}^{\;\;\;h}:=\frac{1}{2\varphi}\/\left( - S_{ij}^{\;\;\;h} + S_{j\;\;\;i}^{\;\;h} - S^h_{\;\;ij}\right)
\end{equation}
\begin{equation}\label{2.9b}
\hat{K}_{ij}^{\;\;\;h} := -\hat{T}_j\delta^h_i + \hat{T}_pg^{ph}g_{ij}
\end{equation}
\begin{equation}\label{2.9c}
\hat{T}_j:=\frac{1}{2\varphi}\/\left( \partial_j\/\varphi + S^{\;\;\;q}_{jq} \right)
\end{equation}
\end{subequations}
By inserting it into eq. (\ref{riemann}), we get the following decomposition of the Ricci tensor
\begin{equation}\label{2.10}
\begin{split}
R_{ij}=\tilde{R}_{ij} + \tilde{\nabla}_j\hat{K}_{hi}^{\;\;\;h} + \tilde{\nabla}_j\hat{S}_{hi}^{\;\;\;h}-\tilde{\nabla}_h\hat{K}_{ji}^{\;\;\;h} - \tilde{\nabla}_h\hat{S}_{ji}^{\;\;\;h} + \hat{K}_{ji}^{\;\;\;p}\hat{K}_{hp}^{\;\;\;h} + \hat{K}_{ji}^{\;\;\;p}\hat{S}_{hp}^{\;\;\;h} \\
+ \hat{S}_{ji}^{\;\;\;p}\hat{K}_{hp}^{\;\;\;h} + \hat{S}_{ji}^{\;\;\;p}\hat{S}_{hp}^{\;\;\;h} - \hat{K}_{hi}^{\;\;\;p}\hat{K}_{jp}^{\;\;\;h} - \hat{K}_{hi}^{\;\;\;p}\hat{S}_{jp}^{\;\;\;h} - \hat{S}_{hi}^{\;\;\;p}\hat{K}_{jp}^{\;\;\;h} - \hat{S}_{hi}^{\;\;\;p}\hat{S}_{jp}^{\;\;\;h}
\end{split}
\end{equation}
inserting (\ref{3.2c}) into the trace of (\ref{3.2a}), we get the relation
\begin{equation}\label{3.5}
-\tilde{R} -\varphi\frac{dV}{d\varphi} + 2V(\varphi) = \Sigma
\end{equation} 
and taking the identities (\ref{2.8}) and (\ref{2.9}) into account, from the decomposition (\ref{2.10}) it is easily seen that the scalar curvatures $\tilde R$ and $R$ satisfy the identity
\begin{equation}\label{3.6}
\tilde{R} = R + \frac{3}{\varphi}\tilde{\nabla}_p\tilde{\nabla}^p\varphi - \frac{3}{2\varphi^2}\partial_p\varphi\partial^p\varphi + \mathcal{H}
\end{equation}
where $\mathcal{H}$ is a function depending on the metric $g_{ij}$, the spin tensor $S_{ij}^{\;\;\;h}$, the scalar field $\varphi$ and their first derivatives: finally, replacing (\ref{3.6}) in (\ref{3.5}) and making use again of (\ref{3.2c}), we end up with an equation of the form
\begin{equation}
\label{kg}
-\tilde{\nabla}_p\tilde{\nabla}^p\varphi+\frac{1}{2\varphi}\partial_p\varphi\partial^p
\varphi+\frac{\varphi[2V-(1+\varphi)\frac{dV}{d\varphi}]} {3}=\frac{\varphi}{3}\left(\Sigma + \mathcal{H}\right)
\end{equation}
as a Klein-Gordon--like equation for the scalar field $\varphi$ which is therefore dynamically coupled to the geometrical background of the theory. The dynamical nature of eq. (\ref{kg}) represents the first important difference discriminating torsional $f(R)$-gravity (\cite{Capozziello:2012gw}, \cite{CCSV1}, \cite{CV4}), where the scalar field $\varphi$ couples {\it algebraically} only to the energy $\Sigma$, and the present hybrid torsional gravity, where the same scalar field $\varphi$ couples {\it dynamically} to the energy $\Sigma$ as well as to the spin through the term $\mathcal{H}$; as it will be pointed out in the next discussion, this point results in a larger dynamical role of the spin with respect to other extended theories of gravity with torsion of least derivative order.

Let us now introduce the tetrad-affine formalism. Considering the Dirac matrices according to the Clifford Algebra and denoted with $\gamma^\mu$, we define $\Gamma^i = e^i_\mu\gamma^\mu$ where $e^\mu_i$ are the tetrad fields associated with the metric $g_{ij}$, and setting $S_{\mu\nu}:=\frac{1}{8}[\gamma_\mu,\gamma_\nu]$, we define the covariant derivative of the Dirac field as $D_i\psi = \de\psi/de{x^i} + \omega_i^{\;\;\mu\nu}S_{\mu\nu}\psi\/$ where $\omega_i^{\;\;\mu\nu}$ is the spin-connection; of course it is again possible to have a metric compatibility condition now given in the form of the vanishing of the covariant derivatives of the tetrads as
\begin{equation}\label{4.2.3}
\Gamma_{ij}^{\;\;\;h} = \omega_{i\;\;\;\nu}^{\;\;\mu}e_\mu^h\/e^\nu_j + e^{h}_{\mu}\partial_{i}e^{\mu}_{j}
\end{equation}
linking the spin-connection to the linear connection defined above.

With these notations, it is possible to define the energy density tensor as
\begin{subequations}
\label{4.2.5}
\begin{equation}
\label{energy}
\Sigma_{ij} := \frac{i}{4}\/\left( \bar\psi\Gamma_{i}{D}_{j}\psi - {D}_{j}\bar{\psi}\Gamma_{i}\psi \right)
\end{equation}
while the spin density tensor is expressed as
\begin{equation}
\label{spin}
S_{ijh}=\frac{i}{2}\bar\psi\left\{\Gamma_{h},S_{ij}\right\}\psi
\equiv\frac{1}{4}\epsilon_{ijhk}\bar{\psi}\gamma^{k}\gamma_{5}\psi
\end{equation}
\end{subequations}
showing the complete antisymmetry for the spin density of the Dirac matter field.

It is easy to prove that the conservation laws \eqref{2.11} are in fact verified by these conserved quantities as soon as the Dirac matter field equations 
\begin{equation}
\label{dirac}
i\Gamma^{h}D_{h}\psi+\frac{i}{2}T_{hj}^{\;\;\;j}\Gamma^h\psi-m\psi=0
\end{equation}
are satisfied by the Dirac matter field, where the explicit contribution of the torsion trace vector is due to the fact that in extended gravity torsion is not completely antisymmetric even for Dirac matter fields, although much can be simplified as we shall show below.

It is now possible to separate all torsional contributions from the purely metric ones, thus substituting torsion with the spin density of the Dirac matter field, then we can express the field eqs. \eqref{3.2a} in the form
\begin{equation}
\label{4.2.9}
\begin{split}
(1+\varphi)\tilde{G}_{ij} = \Sigma_{ij} + \frac{1}{\varphi}\left( - \frac{3}{2}\de\varphi/de{x^i}\de\varphi/de{x^j} + \varphi\tilde{\nabla}_{j}\de\varphi/de{x^i} + \frac{3}{4}\de\varphi/de{x^h}\de\varphi/de{x^k}g^{hk}g_{ij} \right. \\
\left. - \varphi\tilde{\nabla}^h\de\varphi/de{x^h}g_{ij}\right) - \frac{1}{2}V\/(\varphi)g_{ij} + \tilde{\nabla}_h\hat{S}_{ji}^{\;\;\;h} + \hat{S}_{hi}^{\;\;\;p}\hat{S}_{jp}^{\;\;\;h} - \frac{1}{2}\hat{S}_{hq}^{\;\;\;p}\hat{S}_{\;\;p}^{q\;\;\;h}g_{ij}
\end{split}
\end{equation}
where now one has $\hat{S}_{ij}^{\;\;\;h}:=-\frac{1}{2\varphi}S_{ij}^{\;\;\;h}\/$ instead; it is easy to see that equations \eqref{2.11b} are equivalent to the antisymmetric part of the Einstein--like equations \eqref{4.2.9} showing that the significant part of the Einstein--like equations \eqref{4.2.9} is the symmetric one: the latter can be worked out in a similar way as in \cite{FV1} to become
\begin{equation}\label{4.2.15}
\begin{split}
(1+\varphi)\tilde{G}_{ij} = \frac{i}{4}\/\left[ \bar\psi\Gamma_{(i}\tilde{D}_{j)}\psi - \left(\tilde{D}_{(j}\bar\psi\right)\Gamma_{i)}\psi \right]+\frac{3}{64\varphi}(\bar{\psi}\gamma_5\gamma^\tau\psi)(\bar{\psi}\gamma_5\gamma_\tau\psi)g_{ij}+\\
+\frac{1}{\varphi}\left( - \frac{3}{2}\de\varphi/de{x^i}\de\varphi/de{x^j} + \varphi\tilde{\nabla}_{j}\de\varphi/de{x^i}+\frac{3}{4}\de\varphi/de{x^h}\de\varphi/de{x^k}g^{hk}g_{ij}-\varphi\tilde{\nabla}^h\de\varphi/de{x^h}g_{ij}\right)-\frac{1}{2}V\/(\varphi)g_{ij} 
\end{split}
\end{equation}
Also the Dirac equations \eqref{dirac} can be decomposed giving
\begin{equation}\label{4.2.16}
i\Gamma^{h}\tilde{D}_{h}\psi
-\frac{1}{\varphi}\frac{3}{16}\left[(\bar{\psi}\psi)
+i(i\bar{\psi}\gamma_5\psi)\gamma_5\right]\psi-m\psi=0
\end{equation}
in which there is the additional field $\varphi$ playing the role of a running coupling constant for the self-interactions of the Dirac matter field; notice that these interactions have the structure of Nambu-Jona--Lasinio (NJL) potentials. Finally the Klein-Gordon--like equation results to be
\begin{equation}\label{4.2.18}
-\tilde{\nabla}_p\tilde{\nabla}^p\varphi+\frac{1}{2\varphi}\partial_p\varphi\partial^p
\varphi+\frac{\varphi[2V-(1+\varphi)\frac{dV}{d\varphi}]} {3}=\frac{\varphi}{3}\Sigma + \frac{1}{32\varphi}(\bar{\psi}\gamma^\tau\psi)(\bar{\psi}\gamma_\tau\psi)
\end{equation}
describing the dynamics of the scalar field $\varphi$ as it can be easily checked. As we have already noticed, field equation (\ref{kg}) encodes a dynamical evolution for the scalar field $\varphi$ in terms of both the energy $\Sigma$ and the spin, which now can be explicitly written in terms of the four-fermion contribution $\bar{\psi}\gamma^\tau\psi \bar{\psi}\gamma_\tau\psi\geqslant0$ vanishing only in the case one treats single-handed fermions such as neutrino fields in the standard (non-Majorana) representation.

Therefore, hybrid torsional gravity applied to the case of $\frac{1}{2}$-spin fermionic Dirac fields has the specific advantage, compared to any other known field theories, that the Dirac field will be endowed with a non-linear interaction of geometric origin whose coupling displays a running defined in terms of a function, which must be obtained as a solution of a field equation; this field equation for the scalar feels the presence of the spin like, vice versa, the spinor feels the presence of the scalar, in a system of field equations that is tightly coupled. With this toolkit at hand, we may study some of its applications.

As we have said, torsionally-induced gravitational potentials are equivalent to the NJL potentials with a running coupling constant in the Dirac field equation: therefore application of torsion in hybrid gravity for Dirac matter means applying a relativistic analogy of superconductive states where the coupling exhibit a running; the running has to be obtained by solving the Klein-Gordon field equation coming from the hybrid gravity theory. Departures from the standard Einstein gravity signal the beginning of the running coupling effects.

With this scheme in mind, it is easy to search for applications of torsional hybrid gravity to relativistic condensed state physics. In the same spirit in which the NJL model was considered for hadrons in order to provide a description of strong forces before chromodynamics, likewise here we will take into account the case of leptons in order to provide a description of weak forces in alternative to the flavor dynamics of the standard model; another application will be to leptons that are massless and single-handed to provide a description of the neutrino oscillation even for massless states. We will now proceed to study these two models.

\subsection{The case of Weak Forces for Massive Leptons}
In applying the NJL model to the case of leptons, we have to notice that the lack of balance between left-handed and right-handed chiral projections will determine the appearance, beside scalar bosons, also of vector bosons: unlike the NJL model for strong forces, in which only scalars are produced (which can be identified with the pions), for weak forces not only scalar, but also vectors can be produced. They can be identified with Higgs and mediators fields. In this model, however, the Higgs, $W$ and $Z$ particles are composite states of leptons, bound by torsion.

Starting from the field equations for a pair of leptons with torsion interactions
\begin{eqnarray}
&i\gamma^{\mu}\tilde{D}_{\mu}e
-\frac{3}{16\varphi}\overline{e}\gamma_{\mu}e\gamma^{\mu}e
-\frac{3}{16\varphi}\overline{\nu}\gamma_{\mu}\nu\gamma^{\mu}\gamma e-me=0
\label{1}\\
&i\gamma^{\mu}\tilde{D}_{\mu}\nu
-\frac{3}{16\varphi}\overline{e}\gamma_{\mu}\gamma e\gamma^{\mu}\nu=0
\label{2}
\end{eqnarray}
it is a straightforward matter the Fierz re-arrangement, after having introduced, for generality, the parameter $g$ such that the electronic charge can be written as $q=g\sin{\theta}$ in terms of the Weinberg angle and the Yukawa constant $Y$ such that the electronic mass is written as $m=Yv$ in terms of the constant $v=\frac{1}{\sqrt{\sqrt{8}G_{F}}}$. The field eqs. (\ref{1}-\ref{2}) assume the form 
\begin{eqnarray}
\nonumber
&i\gamma^{\mu}\tilde{D}_{\mu}e+\frac{3}{8\varphi}q\tan{\theta}\frac{\cos{\theta}}{g\left[2\sin{\theta}\right)^{2}}
\left[2(\overline{e}_{L}\gamma_{\mu}e_{L}-\overline{\nu}\gamma_{\mu}\nu)
-(2\sin{\theta})^{2}\overline{e}\gamma_{\mu}e\right]\gamma^{\mu}e-\\
\nonumber
&-\frac{3g}{16\varphi\cos{\theta}}\frac{\cos{\theta}}{g\left(2\sin{\theta}\right)^{2}}
\left[2(\overline{e}_{L}\gamma_{\mu}e_{L}-\overline{\nu}\gamma_{\mu}\nu)
-(2\sin{\theta})^{2}\overline{e}\gamma_{\mu}e\right]\gamma^{\mu}e_{L}+\\
\nonumber
&+\frac{3g}{16\varphi}\frac{\left[1-(2\sin{\theta})^{2}\right]}{g\left(2\sin{\theta}\right)^{2}}\left(2\overline{\nu}\gamma_{\mu}e_{L}\right)\gamma^{\mu}\nu-\\
&-Y\frac{3}{16Y\varphi}2(\cos{\theta})^{2}\overline{e}ee
+\frac{3}{8\varphi}(\cos{\theta})^{2}\overline{e}\gamma e\gamma e-me=0
\label{electronic}\\
\nonumber
&i\gamma^{\mu}\tilde{D}_{\mu}\nu+\frac{3g}{16\varphi\cos{\theta}}\frac{\cos{\theta}}{g\left(2\sin{\theta}\right)^{2}}
\left[2(\overline{e}_{L}\gamma_{\mu}e_{L}-\overline{\nu}\gamma_{\mu}\nu)
-(2\sin{\theta})^{2}\overline{e}\gamma_{\mu}e\right]\gamma^{\mu}\nu+\\
&+\frac{3g}{16\varphi}\frac{\left[1-(2\sin{\theta})^{2}\right]}{g\left(2\sin{\theta}\right)^{2}}\left(2\overline{e}_{L}\gamma_{\mu}\nu\right)\gamma^{\mu}e_{L}=0
\label{neutrinic}
\end{eqnarray}
and then
\begin{eqnarray}
\nonumber
&i\gamma^{\mu}\tilde{D}_{\mu}e+\frac{3}{8\varphi}(\cos{\theta})^{2}\overline{e}\gamma e\gamma e
+q\tan{\theta}Z_{\mu}\gamma^{\mu}e-\\
&-\frac{g}{2\cos{\theta}}Z_{\mu}\gamma^{\mu}e_{L}
+\frac{g}{\sqrt{2}}W^{*}_{\mu}\gamma^{\mu}\nu-YHe-me=0
\label{electron}\\
&i\gamma^{\mu}\tilde{D}_{\mu}\nu+\frac{g}{2\cos{\theta}}Z_{\mu}\gamma^{\mu}\nu
+\frac{g}{\sqrt{2}}W_{\mu}\gamma^{\mu}e_{L}=0
\label{neutrino}
\end{eqnarray}
upon definition of the following bosons, being them scalars
\begin{eqnarray}
&H=\frac{3}{8Y\varphi}(\cos{\theta})^{2}\overline{e}e
\label{Higgs}
\end{eqnarray}
or vectors
\begin{eqnarray}
&Z_{\mu}=\frac{3}{8\varphi}\frac{\cos{\theta}}{g\left(2\sin{\theta}\right)^{2}}
\left[2\left(\overline{e}_{L}\gamma_{\mu}e_{L}-\overline{\nu}\gamma_{\mu}\nu\right)
-(2\sin{\theta})^{2}\overline{e}\gamma_{\mu}e\right]
\label{neutral}\\
&W_{\mu}=\frac{3}{8\sqrt{2}\varphi}\frac{\left[1-(2\sin{\theta})^{2}\right]}{g\left(2\sin{\theta}\right)^{2}}
\left(2\overline{e}_{L}\gamma_{\mu}\nu\right)
\label{charged}
\end{eqnarray}
much in the same way that has been followed in \cite{Fabbri:2010ux}; the field equations of the standard model of leptons are given by
\begin{eqnarray}
&i\gamma^{\mu}\tilde{D}_{\mu}e+q\tan{\theta}Z_{\mu}\gamma^{\mu}e
-\frac{g}{2\cos{\theta}}Z_{\mu}\gamma^{\mu}e_{L}
+\frac{g}{\sqrt{2}}W_{\mu}^{\ast}\gamma^{\mu}\nu-YHe-me=0
\label{electronSM}\\
&i\gamma^{\mu}\tilde{D}_{\mu}\nu
+\frac{g}{2\cos{\theta}}Z_{\mu}\gamma^{\mu}\nu
+\frac{g}{\sqrt{2}}W_{\mu}\gamma^{\mu}e_{L}=0
\label{neutrinoSM}
\end{eqnarray}
in terms of the fields $H$ and also $Z_{\mu}$ and $W_{\mu}$ which are, in this case, assumed to be intrinsically structureless: after their comparison, we see that the respective field equations for leptons (\ref{electron}-\ref{neutrino}) and (\ref{electronSM}-\ref{neutrinoSM}) are essentially identical. Here, the scalar field and the leptonic currents (\ref{Higgs}) and (\ref{neutral}-\ref{charged}) play the role of the composite Higgs boson and weak mediators that are fundamental in the standard model. Analogies and differences of the two approaches are the following: here the torsion tensor is justified by the same arguments of generality that in the standard model would justify the introduction of the Higgs potential of spontaneous symmetry breaking, that is in the most general case scalar fields with quadratic and quartic potentials may be added in the same way in which in the torsion tensor is; here the masses of the leptons are justified by the same arguments that in the standard model justify the introduction of the Yukawa potential, that is Yukawa potentials with specific coupling constants are require in the same way in which lepton masses are. However, while the standard model needs to settle two constants for the Higgs potential, the hybrid gravity with torsion needs only a single function $\varphi$ that is automatically determined by means of the Klein-Gordon equation \eqref{kg}. In other words, this approach is more economic and restrictive. Furthermore, the mechanism of spontaneous breaking of the underlying symmetry gives rise to shifts in the vacuum expectation value contributing to the cosmological constant of the universe while, in this approach, no spontaneous symmetry breaking occurs because there is no symmetry.

We would like to specify that in our model, the field configuration has always been asymmetric from the beginning due to the presence of the lepton masses, with no real necessity for the scalar field.

The mediators instead contribute to the dynamics of the leptonic fields considerably through a mechanism of the following type: when two leptons come close one another, their torsional interaction makes them interact binding them together before splitting them apart; in the time, their mutual interaction keeps them as a bound state, they propagate as a single field which may look fundamental if the energy is low enough not to probe its internal compositeness. However, differently from the scalar field, in this case it is possible to compute the partially conserved axial currents (PCAC), which decipher the massiveness of the axial current itself. 

In the case of the standard Einstein theory with torsion, the computation has been done yielding PCAC that closely resemble those of the standard model \cite{Fabbri:2012zd}, but here the presence of the scalar field $\varphi$ changes the computation yielding
\begin{eqnarray}
&\frac{\left[\left(1+\frac{H}{v}\right)\tilde{\nabla}Z^{\mu}+2Z^{\mu}\partial_{\mu}\frac{H}{v}\right]}{\left(1+\frac{H}{v}\right)
\left[\frac{3}{4\varphi}\frac{\cos{\theta}}{g\left(2\sin{\theta}\right)^{2}}
-\frac{1}{m (i\overline{e}\gamma^{5}e)}Z^{\mu}\partial_{\mu}\ln{\left[\frac{1}{\varphi}\left(1+\frac{H}{v}\right)^{2}\right]}\right]}
\!=\!-m i\overline{e}\gamma^{5}e
\label{neutralPCAC}\\
&\frac{\left[\left(1+\frac{H}{v}\right)\left(\tilde{\nabla}_{\mu}W^{\mu}
+iq\tan{\theta}Z^{\mu}W_{\mu}\right)
+2W^{\mu}\partial_{\mu}\frac{H}{v}\right]}
{\left(1+\frac{H}{v}\right)
\left[\frac{3\left[1-(2\sin{\theta})^{2}\right]}{4\sqrt{2}g\varphi}
\left(\frac{3\overline{e}_{L}e_{R}}{16m\varphi}
-\frac{1}{\left(2\sin{\theta}\right)^{2}}\right)
+\frac{1}{m(i\overline{e}\gamma^{5}\nu)}W^{\mu}\partial_{\mu}\ln{\left[\frac{1}{\varphi}\left(1+\frac{H}{v}\right)^{2}\right]}\right]}
\!\!=\!\!mi\overline{e}\gamma^{5}\nu
\label{chargedPCAC}
\end{eqnarray}
for the PCAC; in the standard model, the structureless $Z_{\mu}$ and $W_{\mu}$ would have
\begin{eqnarray}
&\frac{m_{Z}}{\sqrt{\sqrt{2}G_{F}}}\left[\left(1+\frac{H}{v}\right)\!\tilde{\nabla}_{\mu}Z^{\mu}\!
+\!2Z^{\mu}\partial_{\mu}\frac{H}{v}\right]=-mi\overline{e}\gamma^{5}e
\label{constraintneutral}\\
&\frac{m_{W}}{\sqrt{\sqrt{8}G_{F}}}\left[\left(1+\frac{H}{v}\right)\!\left(\tilde{\nabla}_{\mu}W^{\mu}\!+\!iq\tan{\theta}Z^{\mu}W_{\mu}\right)\!+\!2W^{\mu}\partial_{\mu}\frac{H}{v}\right]\!=mi\overline{e}\gamma^{5}\nu
\label{constraintcharged}
\end{eqnarray}
which once again can be compared: it is not difficult to see that these two expressions are equal whenever, in the present approach, the mediators masses display the running
\begin{eqnarray}
&m_{Z}=\frac{\sqrt{\sqrt{2}G_{F}}}{\left(1+\frac{H}{v}\right)
\left[\frac{3}{4\varphi}\frac{\cos{\theta}}{g\left(2\sin{\theta}\right)^{2}}
-\frac{1}{m (i\overline{e}\gamma^{5}e)}Z^{\mu}\partial_{\mu}\ln{\left[\frac{1}{\varphi}\left(1+\frac{H}{v}\right)^{2}\right]}\right]}\\
&m_{W}=\frac{\sqrt{\sqrt{8}G_{F}}}{\left(1+\frac{H}{v}\right)
\left[\frac{3\left[1-(2\sin{\theta})^{2}\right]}{4\sqrt{2}g\varphi}
\left(\frac{3\overline{e}_{L}e_{R}}{16m\varphi}
-\frac{1}{\left(2\sin{\theta}\right)^{2}}\right)
+\frac{1}{m(i\overline{e}\gamma^{5}\nu)}W^{\mu}\partial_{\mu}\ln{\left[\frac{1}{\varphi}\left(1+\frac{H}{v}\right)^{2}\right]}\right]}
\end{eqnarray}
and it is now possible to ask for what conditions these expressions for the masses, in terms of the scalar $\varphi$, would tend to the measured values at the Fermi scale. By following the same footsteps of \cite{Fabbri:2012zd}, it does not take long to acknowledge that this can only happen if the function $\varphi$ is of the order of magnitude of the unity: what this implies is that the hybrid gravity with torsion cannot be the trivial case $\varphi\rightarrow0$ but it must have corrections at least at the Fermi scale. This fact, would however carry the consequence that, at the energy scale at which the composite mediators display their internal compositeness, the corrections to the gravitational effects must appear. So far as our model would say, this should happen not far beyond the TeV energy scale, something that is well within the reach of the present day accelerators like LHC and which could therefore be observed in the near future.

In order to show this, we need to see in what way we can control the magnitude of the function $\varphi$ and this can be done by working out, as last step, possible solutions for the Klein-Gordon eq. \eqref{4.2.18}
\begin{equation}\label{4.2.18.2}
-\tilde{\nabla}_p\tilde{\nabla}^p\varphi+\frac{1}{2\varphi}\partial_p\varphi\partial^p
\varphi+\frac{\varphi[2V-(1+\varphi)\frac{dV}{d\varphi}]} {3}=\frac{\varphi}{3}\Sigma - \frac{1}{32\varphi}(\bar{\psi}\gamma_5\gamma^\tau\psi)(\bar{\psi}\gamma_5\gamma_\tau\psi)
\end{equation}
in the case we are considering, that is in the approximation of weak gravitational field. Let us consider a simple function of the form $f(R)=aR+bR^2$ that we will employ to derive the form of the potential $V(\varphi)=\frac{1}{4b}(\varphi-a)^2$ in the field equation, and for which we will seek for solutions in the form of condensates. Much as in particle physics, this approach it is pursued for the Higgs scalar field, therefore looking for approximately constant solutions: in this case, since $\Sigma=\frac{m}{2}\bar{\psi}\psi$ then we have to solve the field eq. \eqref{4.2.18.2} in the form
\begin{equation}
-\frac{1}{b}v^{2}(v-a)(1+a)=v^{2}m\bar{\psi}\psi+\frac{3}{16}\bar{\psi}\psi\bar{\psi}\psi
\end{equation}
for $\varphi\approx v$ and having taken $\bar{\psi}\gamma^\tau\psi\bar{\psi}\gamma_\tau\psi\approx \bar{\psi}\psi\bar{\psi}\psi$ since $i\bar{\psi}\gamma^5\psi\approx0$ has been assumed for simplicity. Solutions are given as soon as the spinorial contribution is also constant, as
\begin{eqnarray}
&\frac{3}{8}\bar{\psi}\psi\approx v\left[-mv\pm\sqrt{m^{2}v^{2}-\frac{3}{4b}[(v-a)(1+a)]}\right]
\end{eqnarray}
whose validity is subject to the restriction
\begin{eqnarray}
&4m^{2}v^{2}\geqslant\frac{3}{b}[(v-a)(1+a)]
\label{constraint}
\end{eqnarray}
among the parameters, and for large vacuum expectation value compared to the gravitational constant $v\gg a$ and if the other gravitational parameter $b$ is not too small, it is always satisfied. When this condition is verified, then two solutions are possible, corresponding to the two signs of the root: for positive roots, we have that one can write the factor of the non-linear term according to the following expression 
\begin{eqnarray}
&\frac{3\bar{\psi}\psi}{16\varphi}\approx \frac{mv}{2}\left[-1+\sqrt{1-\frac{3}{4bm^{2}v^{2}}[(v-a)(1+a)]}\right]
\end{eqnarray}
which, in the conditions above, can be approximated as 
\begin{eqnarray}
&\frac{3}{16\varphi}\bar{\psi}\psi\approx-\frac{3(1+a)}{16bm}
\end{eqnarray}
while for negative roots
\begin{eqnarray}
&\frac{3\bar{\psi}\psi}{16\varphi}\approx \frac{mv}{2}\left[-1-\sqrt{1-\frac{3}{4bm^{2}v^{2}}[(v-a)(1+a)]}\right]
\end{eqnarray}
in the same approximation, gives
\begin{eqnarray}
&\frac{3}{16\varphi}\bar{\psi}\psi\approx\left(-mv+\frac{3(1+a)}{16bm}\right)
\end{eqnarray}
controlling the magnitude of the interaction. Notice that such a magnitude factor is ruled by the mass $m$ of the particle and the coefficient $b$ of the modified gravity, and that such a modification becomes less and less relevant as long as the particle becomes more and more massive; for light particles, the two contributions tend to become approximately the same, up to the sign, so that they would simply correspond to attractive and repulsive forces, with strength controlled by the factor $\frac{1}{b}$, increasing as $b$ is made smaller and decreasing down to zero for very large $b$, reaching the state of asymptotic gravitational freedom \cite{freedom}. The value of $b$, however, cannot be made small at will, because of the constraint \eqref{constraint}, but of course in the case $v\gg a$ and $b<0$ then \eqref{constraint} is always verified again, and no further constraint is on the magnitude of $b$: by writing $b=-|b|$ with $|b|$ very small we get
\begin{eqnarray}
&\frac{3\bar{\psi}\psi}{16\varphi}\approx\pm\sqrt{\frac{3v(1+a)}{16|b|}}
\end{eqnarray}
again both attractive and repulsive potentials, such that it can be made arbitrarily large as the correction to gravity deciphered by the coefficient $|b|$ decreases. An additional property is that, in this case, we get no dependence on the mass of the particle at all.

We see that as expected, the magnitude of the non-linear interaction can be made arbitrarily strong, without for that spoiling the ultraviolet and infrared limits that are expected, much as what happened in the non-hybrid $f(R)$ gravity with torsion; here however, the function that scales the magnitude of the interaction is obtained as a possible solution, in the given approximations, of a dynamical field equation in which the presence of the spin-spin interactions induced by torsion play a fundamental role.
\subsection{Mixing Oscillations among Massless Neutrinos}
One can deal with neutrino oscillations under the same standards by considering the role of the hybrid-gravity induced scalar field on the coupling. The NJL model for leptons can also be applied to the case in which both leptons are neutrinos, with intriguing consequences: because in this application the chiral symmetry is maximally violated, we would expect no scalar but only vector mediators; in this model therefore, there appears a neutral current that is the mediator of the interactions that produces the neutrino mixing. 

As before, we start from the coupled field equations
\begin{eqnarray}
&i\gamma^{\mu}\tilde{D}_{\mu}\nu_{1}
+\frac{3}{16\varphi}\bar{\nu}_{2}\gamma_{\mu}\nu_{2}\gamma^{\mu}\nu_{1}=0\\
&i\gamma^{\mu}\tilde{D}_{\mu}\nu_{2}
+\frac{3}{16\varphi}\bar{\nu}_{1}\gamma_{\mu}\nu_{1}\gamma^{\mu}\nu_{2}=0
\end{eqnarray}
rearranged as
\begin{eqnarray}
&\!\!\!\!i\gamma^{\mu}\!
\left[\tilde{D}_{\mu}\!\!
\left(\;\!\!\!\!\begin{tabular}{c}$\nu_{1}$\\$\nu_{2}$\end{tabular}\!\!\!\! \;\right)
\!+\!\frac{i}{16\varphi}\!\left(\!\!\!\!\begin{tabular}{cc}
$(\bar{\nu}_{1}\gamma_{\mu}\nu_{1}-\bar{\nu}_{2}\gamma_{\mu}\nu_{2})$ & $2(\bar{\nu}_{1}\gamma_{\mu}\nu_{2})^{\ast}$\\ $2(\bar{\nu}_{1}\gamma_{\mu}\nu_{2})$ & $-(\bar{\nu}_{1}\gamma_{\mu}\nu_{1}-\bar{\nu}_{2}\gamma_{\mu}\nu_{2})$ \end{tabular}\!\!\!\!\right)\!\!
\left(\;\!\!\!\!\begin{tabular}{c}$\nu_{1}$\\ $\nu_{2}$\end{tabular}\!\!\!\!\;\right)\!\right]\!=\!0
\end{eqnarray}
which can be compactified in the form
\begin{eqnarray}
&i\gamma^{\mu}\left[\tilde{D}_{\mu}\nu+ig\vec{A}_{\mu}\cdot\frac{\vec{\sigma}}{2}\nu\right]=0
\label{fieldequationssimplified}
\end{eqnarray}
as soon as the triplet of vectors
\begin{eqnarray}
&\frac{1}{8\varphi}\bar{\nu}\gamma_{\mu}\vec{\sigma}\nu=g\vec{A}_{\mu}
\label{triplet}
\end{eqnarray}
in terms of the constant $g$ and the doublet of spinors
\begin{eqnarray}
&\left(\!\!\begin{tabular}{c}$\nu_{1}$\\$\nu_{2}$\end{tabular}\!\!\right)=\nu
\label{doublet}
\end{eqnarray}
are introduced; notice that the triplet of vectors and the doublet of spinors transform according to the adjoint and fundamental representations of the $SU(2)$ group 
\begin{eqnarray}
&\left[\vec{A}_{\mu}\cdot\frac{\vec{\sigma}}{2}\right]'
=e^{ig\vec{\theta}\cdot\frac{\vec{\sigma}}{2}}
\left[\left(\vec{A}_{\mu}-\partial_{\mu}\vec{\theta}\right)\cdot\frac{\vec{\sigma}}{2}\right]
e^{-ig\vec{\theta}\cdot\frac{\vec{\sigma}}{2}}\\
&\nu'=e^{ig\vec{\theta}\cdot\frac{\vec{\sigma}}{2}}\nu
\end{eqnarray}
implying that there arises a flavour-changing neutral current mixing with a strength proportional to the constant $g$ of the neutrino fields. It is worth noticing that the neutrino oscillations happen precisely because neutrinos are massless \cite{Fabbri:2010hz}. This does not go against any known fact because the allegedly established evidence for neutrino masses comes from the fact that we have observed neutrino oscillations and the \emph{hypothesis} that this can only occur if neutrinos have masses, but that of course does not mean that this hypothesis is true and any alternative model in which neutrinos are massless would do just fine so long as it will be able to produce neutrino oscillations, the only thing we \emph{actually} observed (see \cite{fogli} for a recent review on the status of art).

Notice that that the same arguments outlined above for the PCAC might be reapplied, showing that these flavour-changing neutral currents must be massless.

Notice, in addition, that in the previous case, the function $\varphi$ had to be chosen in order to fit the weak forces at the interaction Fermi scale while here $\varphi$ is chosen to fit the correct neutrino oscillation length: if it would normally be far-fetched to have a single function $\varphi$ interpolating both constants for these two different physical situations, this is certainly not a concern in hybrid torsional-gravity, where the two different values for $\varphi$ simply corresponds to two different gravitational strengths of the two physical systems we are considering. As we move from the TeV energy scale, where gravity might have newly observed effects, to the Solar System scale, where gravity is the observed one, the constant $\varphi$ would gradually move from non-trivial values to increasingly small ones, so that the interaction strength of the composite mediators would correspondingly change from the Fermi scale to the value that is needed to fit the oscillation length.

As a final remark we specify that here too, the same type of solutions for $\varphi$ may be used to control the magnitude of the interaction, and ultimately of the oscillation length.
\section{Discussion and Conclusions}
Several open issues of modern physics are related to the question of interaction couplings that assume given values for specific phenomena; in particular, such couplings can depend on the energy scales also if the underlying interaction is the same. The case of weak forces for leptons and the mixing oscillations are paradigmatic in this picture. A further issue is related to gravitational interaction: it is generally discarded in quantum field theory due to the assumptions that: $i)$ the full quantum gravity regime is very far in energy with respect to the interaction range of other quantum fields and $ii)$ because gravity is fixed to be only the standard GR. Assuming extended theories of gravity could constitute a way out for the above shortcomings, since running coupling constants could be derived.

In this paper, we have considered a background gravitational interaction, the so called Hybrid Gravity, where torsional degrees of freedom are taken into account: the theory consists of a hybrid approach, where standard Einstein gravity is assumed in metric formalism and $f(R)$-gravity added as a correction to GR, is dealt with the metric-affine formalism; here, we considered also torsion into the game with respect to the other cases discussed in literature such as \cite{Harko:2011nh} and \cite{Capozziello:2012ny}. We gave the general field equations for the coupling to Dirac matter fields. When these field equations are decomposed, and torsion substituted with the spin of fermion fields, the resulting matter field equations are found to be those we would have had in the torsionless case but supplemented by self-interactions with a running coupling given in terms of a scalar field. These field equations have been applied to the case of massive leptons in order to describe weak forces and to the case of massless neutrinos in order to describe their oscillations. 

In doing so, we have seen that the presence of torsion in those two physical situations can indeed give rise to corrections that mimics the leptonic weak forces and neutrino oscillations; the different energy scales have been interpolated by the running coupling provided by the Hybrid Gravity. This amounts to unifying both phenomena under the same standards.
\vskip .2in \noindent {\large {\textbf{Acknowledgments}}}\\
S.C. thanks T. Harko, T.S. Koivisto, F.S.N. Lobo and G.J. Olmo for useful discussions and suggestions that allowed to start the topics presented in this paper.


\begin{thebibliography}{50}
\bibitem{w}
S.~Weinberg,
\textit{Phys. Rev.} \textbf{D13}, 974 (1976).
%
\bibitem{p/1}
B.~Pontecorvo,
\textit{Sov. Phys. JETP} \textbf{7}, 172 (1958).
%
\bibitem{p/2}
B.~Pontecorvo,
\textit{Sov. Phys. JETP} \textbf{26}, 984 (1968).
%
\bibitem{g-p}
V.~N.~Gribov and B.~Pontecorvo,
\textit{Phys. Lett. B} \textbf{28}, 493 (1969).
%
\bibitem{a-b}
D.~V.~Ahluwalia and C.~Burgard,
\textit{Gen. Rel. Grav.} \textbf{28}, 1161 (1996).
%
\bibitem{Ahluwalia:2011ea} 
D.~V.~Ahluwalia and S.~P.~Horvath,
Europhys.\ Lett.\  {\bf 95}, 10007 (2011).
%
\bibitem{Ahluwalia:2009rp} 
D.~V.~Ahluwalia and D.~Schritt,
arXiv:0911.2965 [hep-ph].
%
\bibitem{c-f}
C.~Y.~Cardall and G.~M.~Fuller,
\textit{Phys. Rev. D} \textbf{55}, 7960 (1997).
%
\bibitem{p-r-w}
D.~Piriz, M.~Roy and J.~Wudka,
\textit{Phys. Rev. D} \textbf{54}, 1587 (1996).
%
\bibitem{Fabbri:2010ux} 
L.~Fabbri,
\textit{Int. J. Theor. Phys.} \textbf{50}, 3616 (2011).
%
\bibitem{Fabbri:2010hz} 
L.~Fabbri,
\textit{Annales Fond.\ Broglie} \textbf{37}, 33 (2012).
%
\bibitem{Fabbri:2012qr} 
L.~Fabbri and S.~Vignolo,
\textit{Annalen Phys.} {\bf 524}, 826 (2012).
%
\bibitem{Fabbri:2012yg} 
L.~Fabbri and S.~Vignolo,
\textit{Int. J. Theor. Phys.} \textbf{51}, 3186 (2012).
%
\bibitem{Fabbri:2012zd} 
L.~Fabbri,
\textit{Mod. Phys. Lett. A} \textbf{27}, 1250199 (2012).
%
\bibitem{fRgravity} S. Capozziello, Int. J. Mod. Phys. {\bf D 11}, 483 (2002).

\bibitem{fRgravity1}S. Nojiri and S.D. Odintsov, Int.J.Geom.Meth.Mod.Phys. {\bf 4}, 115 (2007).

\bibitem{fRgravity2}S. Capozziello and M. Francaviglia, Gen. Rel. Grav. {\bf 40}, 357 (2008).

\bibitem{fRgravity3}
F.~S.~N.~Lobo, arXiv:0807.1640 [gr-qc].

\bibitem{fRgravity4}
G. J. Olmo, Int. Jou. Mod. Phys. D {\bf 20} 413 (2011).
%
\bibitem{revnoi}
S. Capozziello and M. De Laurentis, Phys. Rept. {\bf 509}, 167 (2011).
%
\bibitem{libro}
S. Capozziello and V. Faraoni, {\it Beyond Einstein gravity: A Survey of gravitational theories for cosmology and astrophysics}, Fundamental Theories of Physics, Vol. 170, Springer, 2010, New York.
%
\bibitem{Capozziello:2012gw}
S.~Capozziello, M.~De Laurentis, L.~Fabbri, S.~Vignolo,
Eur.\ Phys.\ J.\ C {\bf 72}, 1908 (2012) .
%
\bibitem{Harko:2011nh}
T.~Harko, T.~S.~Koivisto, F.~S.~N.~Lobo, G.~J.~Olmo, 
Phys.\ Rev.\ D {\bf 85}, 084016 (2012).
%
\bibitem{Capozziello:2012ny}
S.~Capozziello, T.~Harko, T.~S.~Koivisto, F.~S.~N.~Lobo, G.~J.~Olmo, JCAP {\bf 04}, 011 (2013).
%
\bibitem{b-h-l}
W.~A.~Bardeen, C.~T.~Hill, M.~Lindner,
\textit{Phys. Rev. D} \textbf{41}, 1647 (1990).
%
\bibitem{annalen}
S. Capozziello, M. De Laurentis, Annalen Phys. {\bf 524}, 545 (2012).
%
\bibitem{arturo}
A. Stabile, S. Capozziello, Phys. Rev. D {\bf 87}, 064002 (2013).
%
\bibitem{CCSV1} 
S.~Capozziello, R.~Cianci, C.~Stornaiolo and S.~Vignolo, 
\textit{Class. Quantum Grav.} \textbf{24}, 6417 (2007).
%
\bibitem{CCSV2}
S.~Capozziello, R.~Cianci, C.~Stornaiolo and S.~Vignolo, 
\textit{Int. J. Geom. Meth. Mod. Phys.} \textbf{5}, 765 (2008).
%
\bibitem{CV4} 
S.~Capozziello and S.~Vignolo, 
\textit{Ann. Phys. (Berlin)} \textbf{19}, 238 (2010).
%
\bibitem{Poplawski}
N.~J.~Poplawski, arXiv:0911.0334 [gr-qc].
%
\bibitem{FV1}
L.~Fabbri and S.~Vignolo
\textit{Class. Quantum Grav.} \textbf{28}, 125002 (2011).
%
\bibitem{freedom}
S. Capozziello, R. de Ritis, A.A. Marino, Phys. Lett. A {\bf 249}, 395 (1998).
%
\bibitem{fogli}
G. L. Fogli, E. Lisi, A. Marrone, D. Montanino, A. Palazzo, A. M. Rotunno, 
arXiv:1205.5254 [hep-ph].
\end{thebibliography}
\end{document}